\documentstyle[epsf]{elsart}
\catcode`@=11
\def\chkspace{%
  \relax   
  \begingroup\ifhmode\aftergroup\dochksp@ce\fi\endgroup}
\def\dochksp@ce{%
  \unskip              
  \futurelet\chkspct@k\d@chkspc  
}
\def\d@chkspc{%
  \let\nxtsp@ce=\relax
  \ifx\chkspct@k.\else     
    \ifx\chkspct@k,\else
      \ifx\chkspct@k;\else
        \ifx\chkspct@k!\else
          \ifx\chkspct@k?\else
            \ifx\chkspct@k:\else
              \ifx\chkspct@k)\else
              \ifx\chkspct@k(\else
                \ifx\chkspct@k]\else
                  \ifx\chkspct@k-\else
                    \ifx\chkspct@k\egroup\else  
                      \let\nxtsp@ce=\put@space  
                    \fi
                  \fi
                \fi
              \fi
              \fi
            \fi
          \fi
        \fi
      \fi
    \fi
  \fi
  \nxtsp@ce
}
\def\put@space{$\;$}
\catcode`@=12

\def\ra{\relax\ifmmode \rightarrow\else{{$\rightarrow$}}\fi\chkspace}
\def\etal{{\it et al.}\chkspace}

\def\ie{{\it i.e.}\chkspace}

\def\eg{{\it eg.}\chkspace}

\def\bonafide{{\it bonafide}\chkspace}

\def\ep{{e$^+$e$^-$}\chkspace}

\def\qu{\quad}

\def\gluino{\relax\ifmmode \tilde{g} \else $\tilde{g}$ \fi\chkspace}

\def\qq{\relax\ifmmode q\overline{q}
\else $q\overline{q}$ \fi\chkspace}
\def\ff{\relax\ifmmode f\overline{f}
\else $f\overline{f}$ \fi\chkspace}

\def\bb{\relax\ifmmode b\bar{b}
       \else $b\bar{b}$ \fi\chkspace}
\def\ccrm{\relax\ifmmode {\rm c}\bar{\rm c}
       \else ${\rm c}\bar{\rm c}$ \fi\chkspace}
\def\cc{$c\bar{c}$ \chkspace}
\def\tt{\relax\ifmmode {\rm t}\bar{\rm t}
       \else ${\rm t}\bar{\rm t}$ \fi\chkspace}
\def\ss{\relax\ifmmode {\rm s}\bar{\rm s}
       \else ${\rm s}\bar{\rm s}$ \fi\chkspace}
\def\uu{\relax\ifmmode {\rm u}\bar{\rm u}
       \else ${\rm u}\bar{\rm u}$ \fi\chkspace}
\def\dd{\relax\ifmmode {\rm d}\bar{\rm d}
       \else ${\rm d}\bar{\rm d}$ \fi\chkspace}

\def\qqg{\relax\ifmmode q\overline{q}g
\else $q\overline{q}g$ \fi\chkspace}
\def\bbg{\relax\ifmmode b\overline{b}g
\else $b\overline{b}g$ \fi\chkspace}

\def\afb{\relax\ifmmode A_{FB} \else
{{$A_{FB}$}}\fi\chkspace}
\def\afbb{\relax\ifmmode A_{FB}^b \else
{{$A_{FB}^b$}}\fi\chkspace}
\def\pafb{\relax\ifmmode \tilde{A}_{FB} \else
{{$\tilde{A}_{FB}$}}\fi\chkspace}
\def\pafbb{\relax\ifmmode \tilde{A}_{FB}^b \else
{{$\tilde{A}_{FB}^b$}}\fi\chkspace}

\def\pafbzo{\relax\ifmmode \tilde{A}_{FB}|_{O(0)} \else
{{$\tilde{A}_{FB}|_{O(0)}$}}\fi\chkspace}
\def\pafbfo{\relax\ifmmode \tilde{A}_{FB}|_{\oalp} \else
{{$\tilde{A}_{FB}|_{\oalp}$}}\fi\chkspace}
\def\pafbso{\relax\ifmmode \tilde{A}_{FB}|_{\oalpsq} \else
{{$\tilde{A}_{FB}|_{\oalpsq}$}}\fi\chkspace}
\def\pafbto{\relax\ifmmode \tilde{A}_{FB}|_{\oalpc} \else
{{$\tilde{A}_{FB}|_{\oalpc}$}}\fi\chkspace}

\def\pafbbzo{\relax\ifmmode \tilde{A}_{FB}^b|_{O(0)} \else
{{$\tilde{A}_{FB}^b|_{O(0)}$}}\fi\chkspace}
\def\pafbbfo{\relax\ifmmode \tilde{A}_{FB}^b|_{\oalp} \else
{{$\tilde{A}_{FB}^b|_{\oalp}$}}\fi\chkspace}
\def\pafbbso{\relax\ifmmode \tilde{A}_{FB}^b|_{\oalpsq} \else
{{$\tilde{A}_{FB}^b|_{\oalpsq}$}}\fi\chkspace}
\def\pafbbto{\relax\ifmmode \tilde{A}_{FB}^b|_{\oalpc} \else
{{$\tilde{A}_{FB}^b|_{\oalpc}$}}\fi\chkspace}

\def\afbo0{\tilde{A}_{FB}|_{O(0)}}
\def\afbo1{\tilde{A}_{FB}|_{\oalp}}
\def\afbo2{\tilde{A}_{FB}|_{\oalpsq}}
\def\afbo3{\tilde{A}_{FB}|_{\oalpc}}

\def\lam{\relax\ifmmode \Lambda_{\overline{MS}}
       \else {{$\Lambda_{\overline{MS}}$}}\fi\chkspace}
\def\lamuds{\relax\ifmmode \Lambda^{(3)}_{\overline{MS}}
       \else {{$\Lambda^{(3)}_{\overline{MS}}$}}\fi\chkspace}
\def\lamudsc{\relax\ifmmode \Lambda^{(4)}_{\overline{MS}}
       \else $\Lambda^{(4)}_{\overline{MS}}$\fi\chkspace}
\def\lamudscb{\relax\ifmmode \Lambda^{(5)}_{\overline{MS}}
       \else $\Lambda^{(5)}_{\overline{MS}}$\fi\chkspace}

\def\alp{\relax\ifmmode \alpha_s\else $\alpha_s$\fi\chkspace}
\def\alpbar{\relax\ifmmode \bar{\alpha_s}
       \else $\bar{\alpha_s}$\fi\chkspace}
\def\alpmz{\relax\ifmmode \alpha_s(M_Z)\else $\alpha_s(M_Z)$\fi\chkspace}
\def\alpmzsq{\relax\ifmmode \alpha_s(M_Z^2)
       \else $\alpha_s(M_Z^2)$\fi\chkspace}

\def\oalp{\relax\ifmmode O(\alpha_s)\else{{O($\alpha_s$)}}\fi\chkspace}
\def\oalpsq{\relax\ifmmode O(\alpha_s^2)
           \else{{O($\alpha_s^2$)}}\fi\chkspace}
\def\oalpc{\relax\ifmmode O(\alpha_s^3)
           \else{{O($\alpha_s^3$)}}\fi\chkspace}
\def\oalpf{\relax\ifmmode O(\alpha_s^4)
           \else{{O($\alpha_s^4$)}}\fi\chkspace}

\def\rb{\relax\ifmmode R_3^b/R_3^{all}
           \else{{$R_3^b/R_3^{all}$}}\fi\chkspace}
\def\rc{\relax\ifmmode R_3^c/R_3^{all}
           \else{{$R_3^c/R_3^{all}$}}\fi\chkspace}
\def\ruds{\relax\ifmmode R_3^{uds}/R_3^{all}
           \else{{$R_3^{uds}/R_3^{all}$}}\fi\chkspace}
\def\ri{\relax\ifmmode R_3^i/R_3^{all}
           \else{{$R_3^i/R_3^{all}$}}\fi\chkspace}
\def\rj{\relax\ifmmode R_3^j/R_3^{all}
           \else{{$R_3^j/R_3^{all}$}}\fi\chkspace}
\def\alpi{\relax\ifmmode \alpha^i_s/\alpha^{all}_s
           \else{{$\alpha^i_s/\alpha^{all}_s$}}\fi\chkspace}

\def\mbz{\relax\ifmmode m_b(M_Z)
           \else{{$m_b(M_Z)$}}\fi\chkspace}
\def\mbb{\relax\ifmmode m_b(M_b)
           \else{{$m_b(M_b)$}}\fi\chkspace}

\def\z0{{$Z^0$}\chkspace}
\def\z0{\relax\ifmmode Z^0 \else {$Z^0$} \fi\chkspace}
\def\Dst{\relax\ifmmode {\rm D}^* \else {D$^*$}\fi\chkspace}
\def\Dpl{\relax\ifmmode {\rm D}^+ \else {D$^+$}\fi\chkspace}
\def\D0{\relax\ifmmode {\rm D}^0 \else {D$^0$}\fi\chkspace}
\def\Kst{\relax\ifmmode {\rm K}^* \else {K$^*$}\fi\chkspace}
\def\K0{\relax\ifmmode {\rm K}^0_s \else {K$^0_s$}\fi\chkspace}
\def\Kpl{\relax\ifmmode {\rm K}^+ \else {K$^+$}\fi\chkspace}
\def\Kstz{\relax\ifmmode {\rm K}^{*0} \else {K$^{*0}$}\fi\chkspace}

\def\beq{\begin{equation}}
\def\eeq{\end{equation}}
\def\bea{\begin{eqnarray}}
\def\eea{\end{eqnarray}}

\begin{document}

\hfill{OUNP-99-10}

\hfill{August 2 1999}

\vskip .5truecm

\begin{frontmatter}

\title{A CCD Vertex Detector for a High-Energy Linear \ep Collider}

\author{P.N. Burrows$^1$}

\address{Particle \& Nuclear Physics, Keble Rd., Oxford, OX1 3RH, UK.}

\thanks{Supported by the UK Particle Physics \& Astronomy Research Council\\
email: {\it p.burrows1@physics.ox.ac.uk}}

\begin{abstract}
I present a summary of the experience with CCD-based vertex detectors at
the SLD experiment at SLAC, and discuss their advantages for use at
a future high-energy linear \ep collider. The extensive R\&D programme
to improve further the vertexing capabilities of CCD detectors is also
outlined.

\end{abstract}

\end{frontmatter}

\vskip 1truecm

\centerline{\it Presented at Vertex99, Texel}
\centerline{\it The Netherlands, 20-25 June 1999}

\vskip 1.2truecm

\section{Introduction to CCDs}

Charge-coupled devices (CCDs) were originally applied in high-energy particle 
physics at a fixed-target charm-production experiment, and their utility
for high-precision vertexing of short-lived particles was quickly 
realised~\cite{accmor}. More recently two generations of CCD vertex detectors
(VXDs) were used in the \ep colliding-beam environment of the SLD experiment
at the first linear collider, SLC, at SLAC.

CCDs are silicon pixel devices which are widely used for imaging; one common
application is in home video cameras, and there is extensive industrial
manufacturing experience in Europe, Japan and the US. 
CCDs can be made with high pixel granularity. For example, those used at SLD
comprise 20$\times$20 $\mu$m$^2$ pixels, offering the possibility of intrinsic
space-point resolution of better than 4 $\mu$m, determined from the centroid of 
the small
number of pixels which are hit when a charged particle traverses the device.
The active depth in the silicon is only 20 $\mu$m, so each pixel is
effectively a cube of side 20 $\mu$m, yielding true 3-dimensional spatial
information. Furthermore, this small active depth allows CCDs to be made
very thin, ultimately perhaps as thin as 20 $\mu$m, which corresponds to less
than 0.1\% of a radiation length ($X_0$), and yields a 
very small multiple scattering of charged particles. 
Also, large-area CCDs can be made for scientific purposes, allowing an
elegant VXD geometry with very few (if any) cracks or gaps for readout
cables or support structures. For example, the second-generation CCDs
used at SLD were of size 80 $\times$ 16 mm$^2$.

The combination of superb spatial resolution, low multiple scattering,
and large-area devices, with a decade of operating experience
at the first linear collider, SLC, hence makes CCDs a very attractive
option for use in a vertex detector at a second-generation linear
collider (LC). Such colliders are being actively pursued by
consortia centred around SLAC and KEK (NLC/JLC), DESY (TESLA)
and CERN (CLIC).

In Section 2 I review briefly the SLD CCD VXD experience; more details
are given in a complementary presentation~\cite{toshi}. In Section 3
I discuss the physics requirements for the LC VXD.
In Section 4 I
present the current conceptual design, and the simulated flavour-tagging
performance. The R\&D programme that is underway to
achieve these goals is described in Section 5.
In Section 6 I give a brief summary and outlook.

\section{SLD CCD VXD Experience}

\begin{figure}[ht]
 \hspace*{0.5cm}
   \epsfxsize=5.625in
   \epsfysize=5in
   \begin{center}\mbox{\epsffile{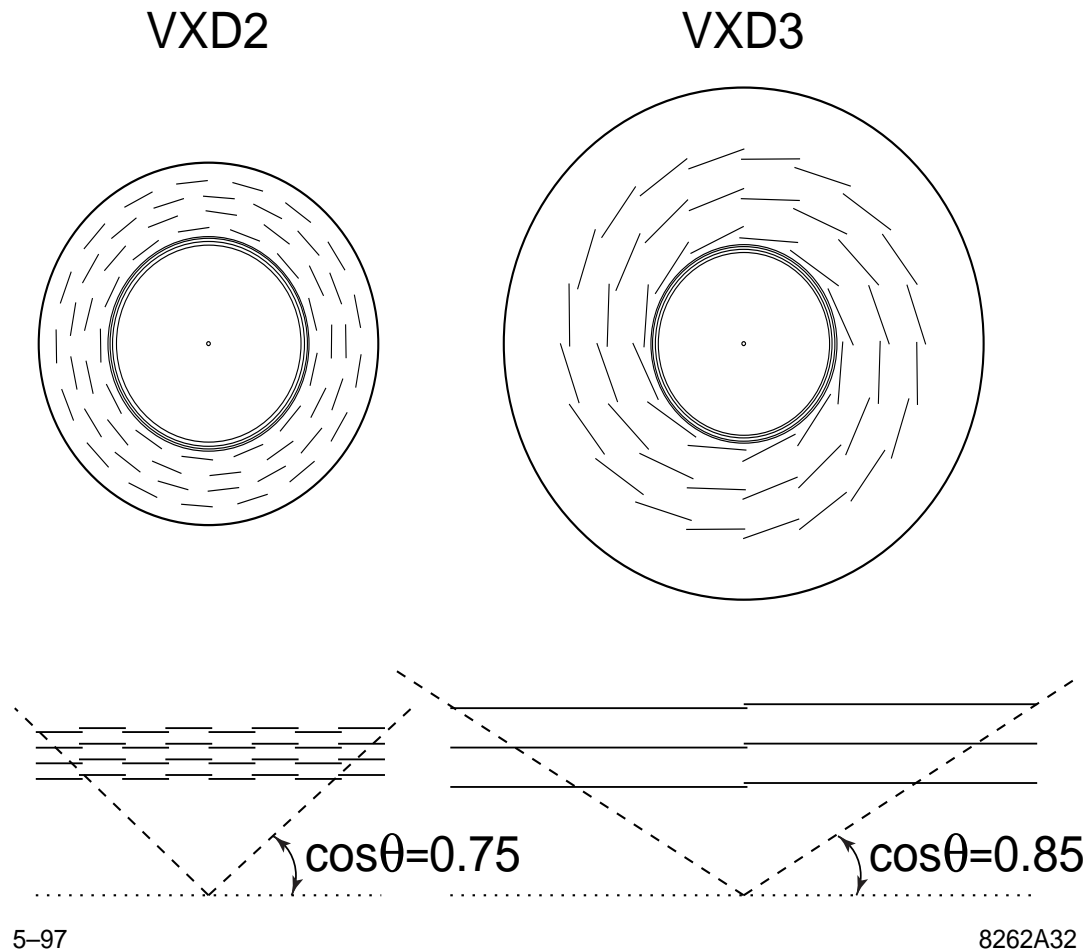}}\end{center}
  \caption{
Comparison of VXD2 and VXD3 geometries
    }
 \label{figvxd23}
\end{figure}

The SLD experiment has utilised three CCD arrays for heavy-flavour
tagging in \z0 decays. In 1991 a 3-ladder prototype detector, VXD1,
was installed for initial operating experience. In 1992 a complete
four-layer vertex detector, VXD2, was installed and operated until 1995. 
VXD2~\cite{vxd2} utilised 64 ladders arranged in 4 incomplete azimuthal
layers (Fig.~\ref{figvxd23}). 
Due to the incomplete coverage a track at a polar angle of
90$^{\circ}$ to the beamline passed through, on average, only 2.3
layers, and $\geq$2-hit coverage extended down to polar angles within
$|\cos\theta|\leq0.75$. The device contained a total of 512 roughly
$1\times1$cm$^2$ CCDs, giving a total of 120M pixels.

In 1996 a brand new detector, VXD3~\cite{vxd3}, was installed that capitalised
on improvements in CCD technology since VXD2 was originally designed.
The main improvement was to utilise much larger, $8\times1.6$cm$^2$,
and thinner ($\times$ 3)  CCDs,
which allowed a significantly improved geometry (Fig~\ref{figvxd23}).
Ladders were formed from two CCDs placed end-to-end (with a small
overlap in coverage) on opposite sides of a beryllium support beam,
and arranged in 3 complete azimuthal layers, with a `shingled'
geometry to ensure no gaps in azimuth. A much better acceptance of
$|\cos\theta|\leq 0.85$ ($\geq$3 hits) and
$|\cos\theta|\leq 0.90$ ($\geq$2 hits) was achieved with these
longer ladders. 96 CCDS were used, giving a total count
of 307M pixels.

In operations from 1996 through 1998 VXD3 performed beautifully,
yielding a measured single-hit resolution of 3.8 $\mu$m, and a track
impact-parameter resolution of 7.8 $\mu$m (9.7 $\mu$m) in $r-\phi$
($r-z$) respectively, measured using 46 GeV $\mu$ tracks in \z0 \ra
$\mu^+\mu^-$ events. The multiple scattering term was found to be 
$33/p\sin^{3/2}\theta$ $\mu$m. The measured precision on the
position of the micron-sized (mm-long) SLC interaction point (IP) was found
to be 3 $\mu$m in $r-\phi$ (30 $\mu$m in $r-z$) respectively.
As an illustration of the vertexing performance, the resolution on 
the decay-length of $B_s$ mesons w.r.t. the IP was estimated to be 
characterised by a double-Gaussian function with widths of 46 $\mu$m,
representing 60\% of the population, and 158$\mu$m representing the
remainder, which is outstanding compared with other \ep experiments.
For inclusive $b$-hemisphere tagging a sample purity of 98\% can be
obtained with a tag efficiency of up to 45\%, and for inclusive 
$c$-tagging a sample purity of around 70\% can be obtained with a
tag efficiency of up to 20\%. Again, this performance is
unsurpassed by other experiments.

It is worth noting several important lessons learned from the
SLD VXD3 experience:

\begin{itemize}
 
\item
The intrinsic spatial resolution of 6 $\mu$m, which one would
naively estimate for a pixel size of 20 $\mu$m, was significantly
improved to better than 4 $\mu$m by capitalising on the charge-sharing
between several pixels and performing cluster centroid-finding.

\item
The complete 3-layer coverage for $|\cos\theta|\leq 0.85$ 
allowed vector hits to be found within the VXD alone, which
could then be included at an earlier stage of the track-finding
algorithm, allowing a `pointing out' rather than `pointing in'
approach to track-linking with the main tracking chamber.
One (or two) VXD hits were also included in the track-finding
of low polar-angle $\mu^+\mu^-$ events, yielding an improved
acceptance and significantly better momentum determination.

\item
Even though in each triggered event there were, on average, roughly 
15,000 pixels hit by background particles from SLC,
the large pixellation of VXD3 ensured that the occupancy 
was approximately $5\times10^{-5}$, yielding essentially zero
confusion between \bonafide hits on tracks and background hits.

\item
The long readout time of VXD3, around 180 ms, did not lead to
any deadtime at all. If a second trigger was taken during the
readout period of the previously triggered event, the VXD readout
was simply restarted and two `overlapping frames' were recorded.
The low hit density ensured that there was zero
confusion of hits between the two overlapping events.

\item
Despite the VXD3 ladders being very thin, roughly 0.4\% $X_0$,
multiple scattering dominated the impact-parameter resolution
for tracks with momenta less than 3 GeV/$c$, \ie the vast
majority of tracks in \z0 decays.

\end{itemize}

These lessons have proven invaluable for consideration of the
design of a VXD for the future LC.

\section{Linear Collider Physics Demands}

The second-generation linear collider will probably be built to operate
at c.m. energies in the range between the current LEP2
energy of around 200 GeV and up to around 0.8 - 1 TeV. The strategy
for choosing the energy steps will be developed as we learn more about 
the Higgs boson(s) and beyond-Standard-Model particles from searches
at LEP2, the Tevatron, HERA and the LHC.
Consideration of a high-statistics run at the \z0 resonance, for
super-precise measurements of electroweak parameters, is
also under discussion.

Many of the interesting physics processes can be
characterised as multijet final states containing heavy-flavour
jets. Some representative examples are:


1) \ep $\qu$ \ra $\qu$ \z0 $H^0$ $\qu$ \ra $\qu$ \qq \bb

2) \ep $\qu$ \ra $\qu$ \z0 $H^0$ $\qu$ \ra $\qu$ \qq \cc

3) \ep $\qu$ \ra $\qu$ \z0 $H^0$ $\qu$ \ra $\qu$ \qq $\tau^+\tau^-$ 

4) \ep $\qu$ \ra $\qu$ \tt $\;\;\qu\qu\qu$  \ra$\qu$ $b W^+$ $\bar{b} W^-$

5) \ep $\qu$ \ra $\qu$ $H^0$ $A^0$ $\qu\;\;$ \ra$\qu$ \tt \tt

6) \ep $\qu$ \ra $\qu$ $\tilde{t}\,\tilde{\bar{t}}$ $\qu\qu\qu\;\;$ \ra$\qu$ 
$\tilde{\chi^0}\,c\,\tilde{\chi^0}\, \bar{c} $

7) \ep $\qu$ \ra $\qu$ \tt $H^0$  $\qu\qu\;$ \ra$\qu$ $b W^+$ $\bar{b} W^-$ \bb

It should be noted that charm- and $\tau$-tagging, as well as $b$-tagging,
will be very important. For example, measurements of the branching ratios
for (the) Higgs boson(s) to decay into $b$, $c$, and $\tau$ pairs (examples 1-3) 
(and/or $W$, \z0 and $t$ pairs for a heavy Higgs)
will be
crucial to map out the mass-dependence of the Higgs coupling and to
determine the nature (SM, MSSM, SUGRA $\ldots$) of the Higgs particle(s).

Example 4 could yield a 6-jet final state containing 2 $b$-jets.
Example 5 could yield a 12-jet final state containing 4 $b$-jets.
Example 6 comprises 2 charm jets + missing energy in the final state.
Example 7 could yield an 8-jet final state containing 4 $b$-jets.

Because of this multijet structure, even at $\sqrt{s}$ = 1 TeV many of these
processes will have jet energies in the range 50 \ra 200 GeV, which is not
significantly larger than at SLC, LEP or LEP2. The track momenta will be
correspondingly low. For example, at $\sqrt{s}$ = 500 GeV the mean track
momentum in \ep \ra \qq events is expected to be around 2 GeV/$c$, so
that with the current SLD VXD3 multiple scattering would 
limit the impact-parameter resolution for the majority of tracks! 

Furthermore, some of these processes may lie close to
the boundary of the accessible phase space, suggesting that extremely high
flavour-tagging efficiency will be crucial for identifying a potentially small
sample of events above a large multijet combinatorial background.
It is worth bearing in mind that a doubling of the single-jet
tagging efficiency at high
purity is equivalent to a luminosity gain of a factor of 16 for a 4-jet
tag (examples 5, 7); it is likely to be a lot cheaper (and easier) 
to achieve this gain by
building a superior VXD than by increasing the luminosity of the
accelerator by over an order of magnitude!
 
\section{LC VXD Conceptual Design}

\begin{figure}[t]
 \hspace*{0.5cm}
\vspace*{-6cm}
   \epsfxsize=5.625in
   \epsfysize=7.5in
   \begin{center}\mbox{\epsffile{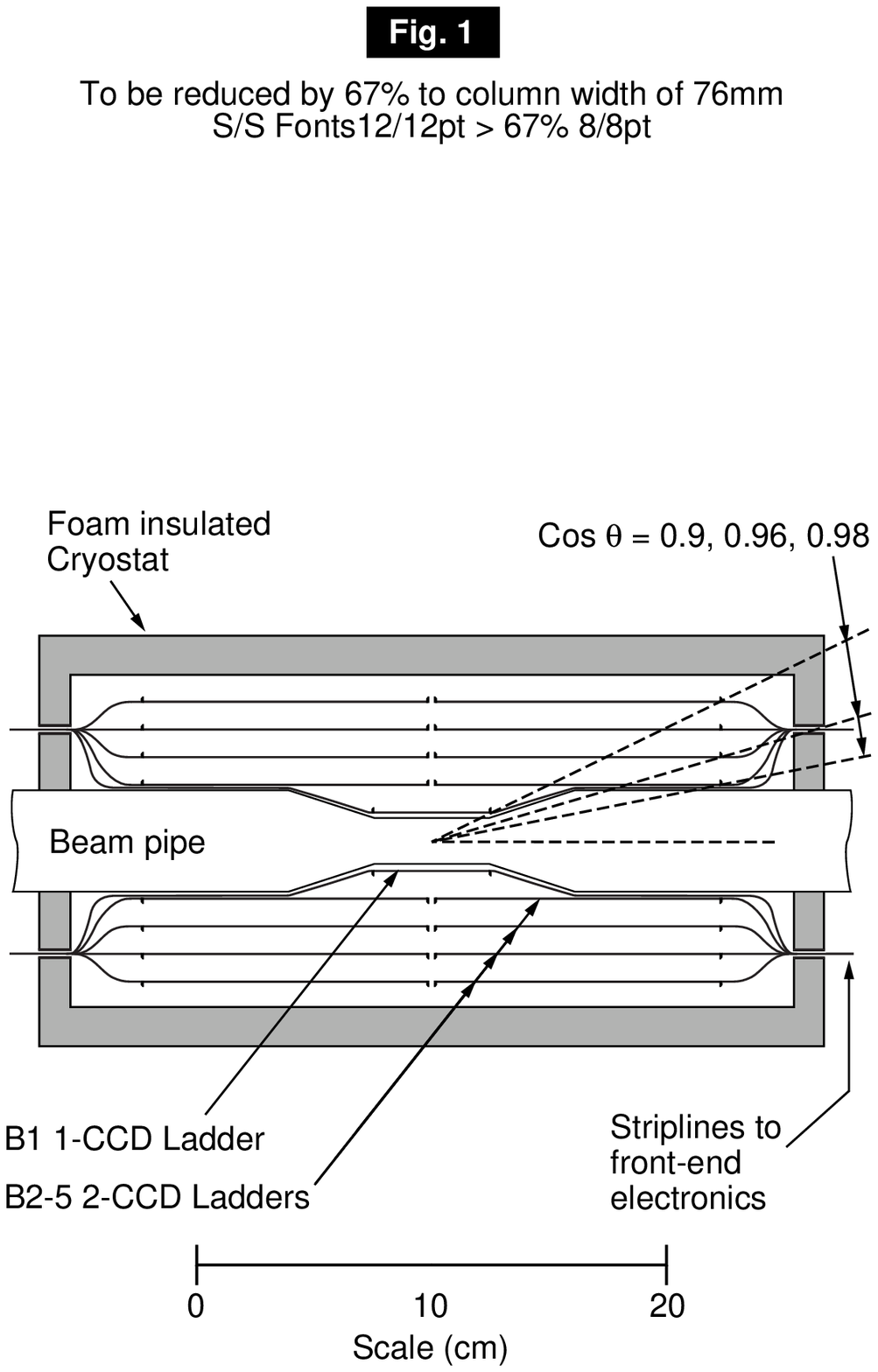}}\end{center}
  \caption{
Conceptual design of the LC VXD
    }
\label{figdesign}
\end{figure}

The LC VXD conceptual design is illustrated in Fig.~\ref{figdesign}.
The main goals to be met to achieve this design can be summarised as:

\begin{itemize}

\item
Utilise large-area CCDs to construct a geometrically elegant, large array.

\item
Obtain VXD self-tracking with redundancy by building a 5-layer device.

\item
Require as short a track extrapolation to the IP as possible by 
putting the first layer as close as 12mm to the beamline.

\item
Reduce multiple scattering by thinning the ladders to as little
as 0.1\% $X_0$ per layer.

\item
Maintain the low occupancy, and hence zero hit confusion, by
increasing the pixel readout rate to 50 MHz.

\item
Improve the radiation tolerance beyond the $10^{10}$ neutrons/cm$^2$
level.

\end{itemize}

\begin{figure}[ht]
 \hspace*{0.5cm}
   \epsfxsize=5in
   \epsfysize=7.5in
   \begin{center}\mbox{\epsffile{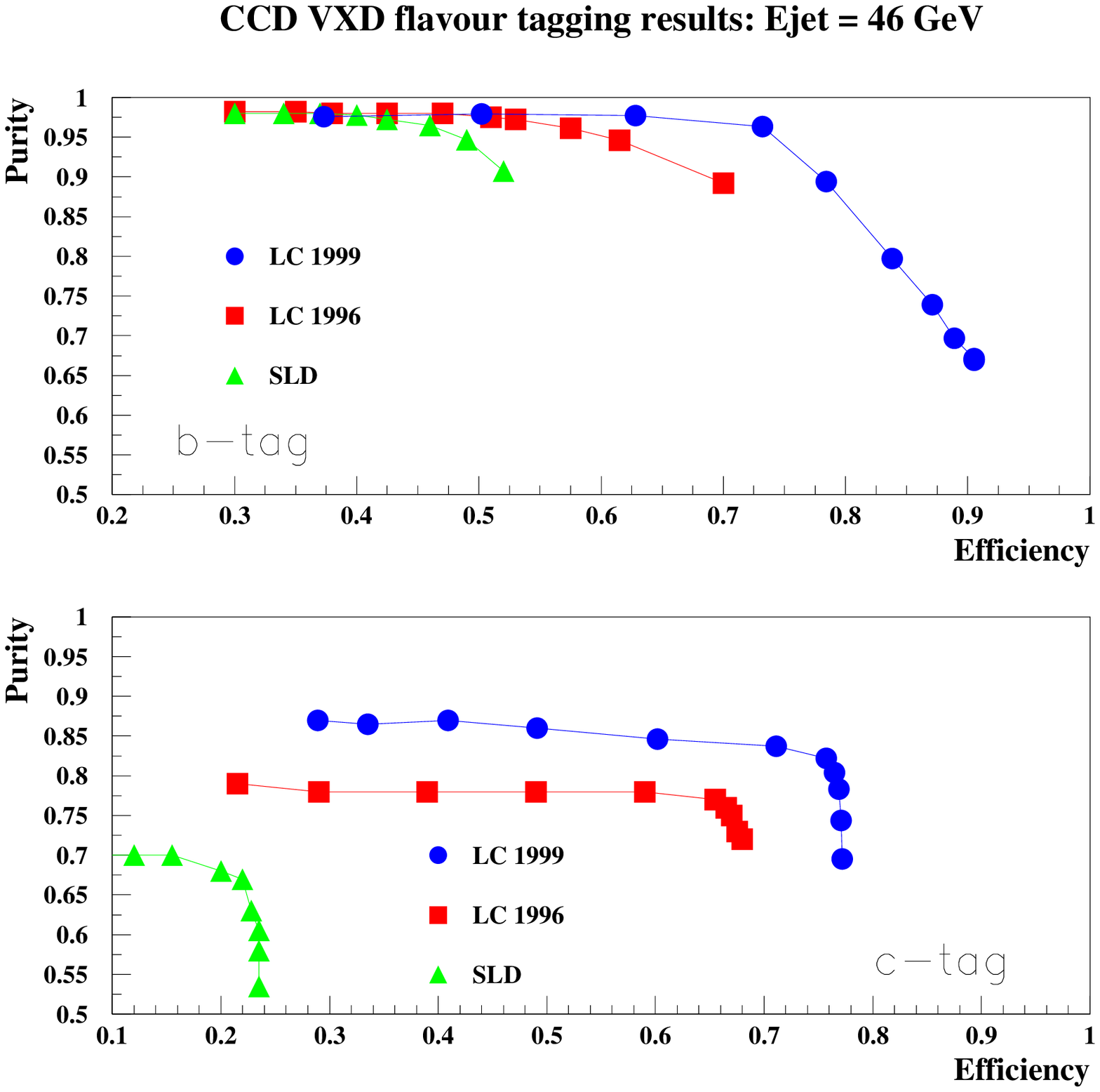}}\end{center}
\vspace*{-2cm}
  \caption{
Purity vs. efficiency achievable for a) inclusive $b$-jet tagging, (b)
inclusive $c$-jet tagging, for 46 GeV jets and a \z0 flavour mix.
    }
\label{figeff}
\end{figure}

We have simulated the jet flavour-tagging performance that could be achieved
with such a `dream' VXD. We have adapted the SLD charm- and bottom-jet
tags which are based on the mass of secondary decay vertices~\cite{sldtag}
reconstructed using a topological vertex-finding algorithm~\cite{dave}.
The purity vs. efficiency trajectories are shown in Fig.~\ref{figeff},
where the current SLD results, as well as results for an earlier LC VXD
design~\cite{snowmass}, are shown for reference. For $b$-jet tagging
a sample purity of 98\% can be maintained for a tagging efficiency up to
around 70\%, almost a factor of two better than the current (world's best)
SLD VXD3. In the case of $c$-jet tagging a sample purity of 85\% can be
achieved for an efficiency up to 75\%, which is a substantial gain in
both purity and efficiency ($\times$ 3.5) w.r.t. SLD VXD3.

A substantial R\&D programme is needed to achieve this 
impressive flavour-tagging potential.  

\section{R\&D Programme}

The UK-based Linear Collider Flavour Identification collaboration
was formed, and was
approved by the UK funding committee in October 1998 to initiate 
a 3-year programme of research and development~\cite{lcfi} in order
to address these design challenges. The collaboration is working closely
with the UK-based CCD manufacturer, EEV, as well as with colleagues in
the US and Japan who are also engaged in CCD R\&D for the LC VXD.

Table~\ref{factors} summarises the improvement factors that it is
hoped to achieve, relative to the current SLD VXD3, for various
parameters.

\begin{table}[ht]
\begin{center}
\begin{tabular}{|l|c|c|c|}
\hline
Item &             SLD & LC  & factor \\ \hline
 & & &  \\
longest CCD (mm) & 80  & 125 &   1.6 \\  
largest CCD area (mm$^2$) &  1280  &  3000 &  2.3 \\  
ladder thickness (\% $X_0$) $\qu$ &  0.4 &  0.12 &  3.3 \\  
layer 1 radius (mm) &  28  & 12 &  2.3 \\  
readout rate (MHz)  & 5  & 50 &   10 \\  
\# ladders          & 48  &  64 &   1.3 \\  
\# pixels (M)       & 307  & 700 &   2.3 \\  \hline
\end{tabular}
\vspace*{0.5cm}
\caption{CCD performance improvement factors required for the LC VXD}
\vspace*{0.5cm}
\label{factors}
\end{center}
\end{table}

In the first phase of the R\&D programme so-called `setup grade' CCDs
will be purchased from EEV and used to test individual design aspects.
These are typically devices with some defect(s) of a mechanical or
electrical nature, but which are perfectly adequate for testing
unaffected performance aspects. Two modular CCD test setups are
being constructed, one located at the Rutherford-Appleton Laboratory (RAL)
and the other at Liverpool University. The aim is to use the two
test rigs to focus on
complementary aspects: readout and electrical tests at RAL, and radiation
damage studies and low-temperature operation at Liverpool. In addition, 
a metrology setup for mechanical testing and measurement is being
developed at Oxford and RAL, which will focus on thin prototype
ladder supports and thermal distortion characterisation.

In a later phase custom-made CCDs may be commissioned from the
manufacturer, and system design/integration issues will be
investigated. 

By designing modular test setups in which only the local motherboard
is CCD-specific it is hoped that CCDs can be readily exchanged with
our US and Japanese colleagues, as well as with different CCD
manufacturers.

\subsection{CCD Area}

A modest increase in length of the longest CCD, by a factor of 1.6,
is needed for the geometry shown in Fig.~\ref{figdesign}, with a
corresponding area increase by just over a factor of two.
With the widespread movement to 8-, and even 12-, inch wafers 
in the silicon chip industry there is not believed to be any
fundamental obstacle towards manufacturing CCDs of this size.
However, the increased area does impinge directly on 
the rigidity/stability requirements for the low-mass support structures.

\subsection{Ladder Thickness}

If the CCDs can be successfully thinned down to the thickness of the
epitaxial active silicon surface layer, they could in principle
be as thin as 20 $\mu$m, or 0.02\% $X_0$. Such CCDs would, if 
unsupported, immediately curl up into a `swiss roll', and so the
thinning process, and adhesion to a thin (beryllium) support beam,
must be carefully thought out. 

\begin{figure}[ht]
 \hspace*{0.5cm}
   \epsfxsize=5.625in
   \epsfysize=7.5in
   \begin{center}\mbox{\epsffile{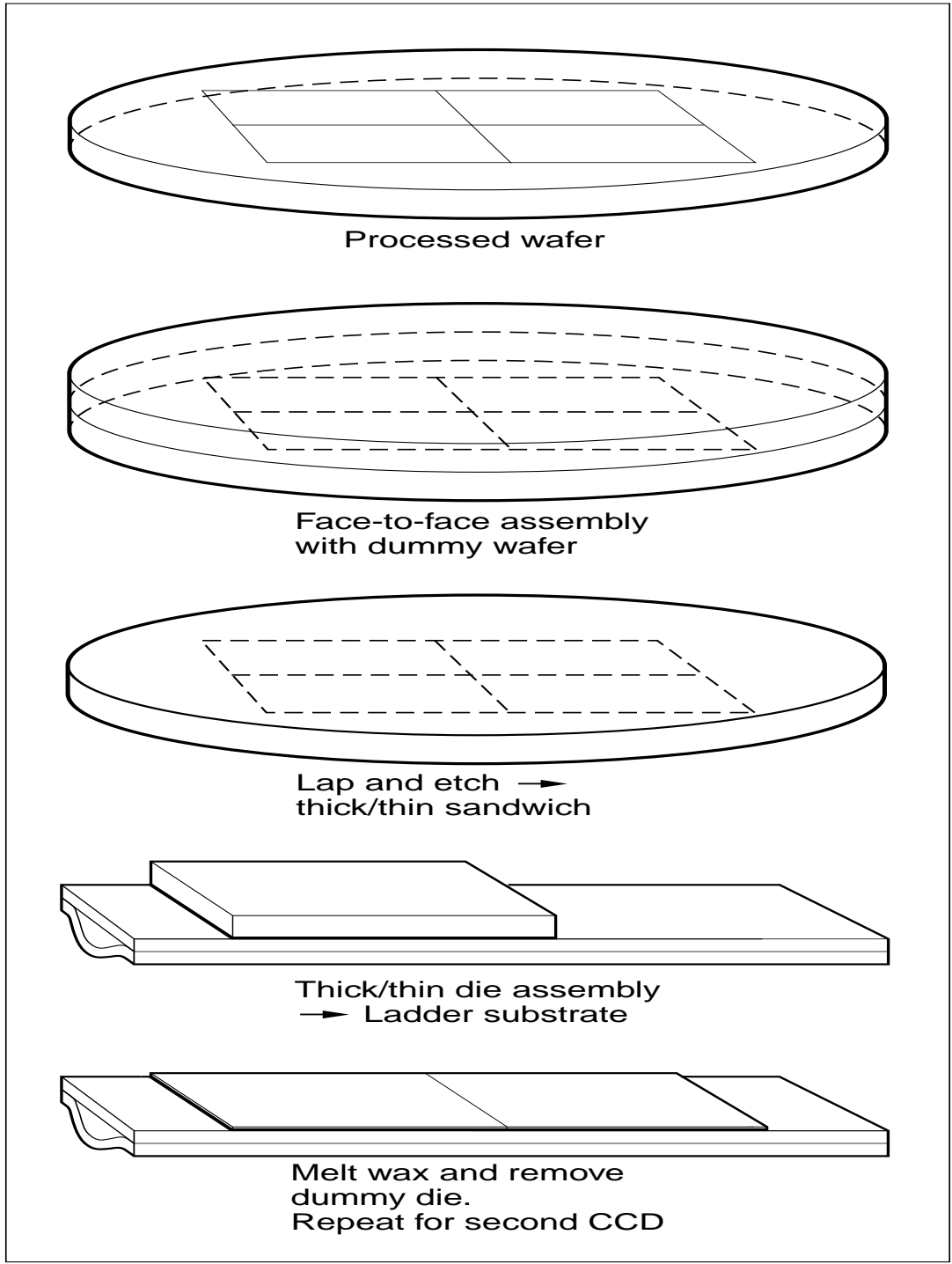}}\end{center}
  \caption{
Hypothetical sequence of operations in thinning the CCDs and attaching
them to low-mass support beams.
    }
 \label{figthin}
\end{figure}

A schematic thinning process has been devised and is illustrated in
Fig.~\ref{figthin}. Before thinning one would adhere the
top surface of a processed CCD wafer 
with temporary adhesive, \eg wax, to a dummy wafer. 
One would then lap and etch the back of the processed wafer to the desired 
thickness, and dice the wafer into individual CCD units.
One would then adhere the back CCD surface to the support beam with
permanent adhesive, and finally disengage the dummy silicon block by
removing the temporary adhesive. 

The support beam itself requires careful design to achieve a low-mass
structure with the desired planarity and mechanical stability. One
possibility is to use a thin flat Be beam with an intrinsic `omega' or `V'
support structure. Finite-element analysis simulations have shown that
such structures offer the possibility of small, and predictable,
deformations under temperature cycling of the order of tens of $\mu$m.

If the support beam comprises 250 $\mu$m Be-equivalent, or 0.07\% $X_0$,
and the adhesive an additional 0.02\% $X_0$, the total ladder material
budget might be made as low as 0.11\% $X_0$.

\subsection{Backgrounds, Inner Radius and Readout Rate} 

The radial position of the innermost layer w.r.t. the beamline is
strongly influenced by the accelerator-related backgrounds, and is
correlated with the pixel readout rate, which determines the 
hit density accumulated during the CCD readout cycle, and hence the
degree of fake hit confusion for \bonafide tracks.

The main sources of accelerator-related backgrounds are

\begin{itemize}

\item
Muons from beam interactions with upstream collimators.

\item
\ep pairs from converted photons and `beamstrahlung'.

\item
Photoproduced neutrons from the interaction region
material and back-shine from the beam dumps.

\item
Hadrons from beam-gas and $\gamma\gamma$ interactions.

\end{itemize}

From the occupancy point-of-view the most serious are the
\ep pairs. For example, beam-beam interaction simulations~\cite{napoly}
indicate that tens of thousands of \ep pairs will be created {\it per
bunch crossing} of the accelerator. A significant fraction of these
populate the $>$ 1 GeV tail of the energy distribution, and it is clear that
a large detector magnetic field will be required to contain the bulk
within the beampipe, and maintain an acceptably low hit density
in the VXD. Field strengths of between 3 and 6 T are being considered
by the detector working groups. For example, in a 3 T field at TESLA
0.2 hits/mm$^2$ {\it per beam-crossing}
are expected at the nominal first-layer radius of 12 mm.  
At NLC/JLC the corresponding figure
is 0.1 hits/mm$^2$ in a 6 T field. 

At first sight these numbers do not appear forbidding. However, one must
then consider the serial pixel readout of the CCD and the 
time-structure of the accelerator bunch trains.
In the NLC/JLC case
there are 100 bunches per train, with a bunch separation of 1.4 ns and
a train separation of 8.3 ms. Hence if the CCD pixel readout rate 
is 5 MHz, as in SLD VXD3, it would take roughly 20 bunch-trains to
read out a complete CCD, implying an integrated hit density of
200/mm$^2$. Since there are 2500 pixels/mm$^2$ this implies an
occupancy of almost 10\%, which would lead to significant hit confusion.
An occupancy of 1\% would be much more manageable, and this could
be achieved by increasing the pixel readout rate by a factor of 10, 
to 50 MHz. This will be one of the main topics in the CCD R\&D
programme. 

In the TESLA design the situation is made more difficult by the
fact that there are roughly 3000 bunches per train, with a bunch
separation of 337 ns and a train
separation of 200 ms. Although a CCD could be read out completely, even at
5 MHz, between bunch trains, the hit density integrated during the
readout cycle would be 600/mm$^2$. Even with an increased pixel readout 
rate of 50 MHz the resulting 60 hits/mm$^2$ is not comfortable.
For this reason we are also investigating the possibility of a 
higher-multiplex CCD readout scheme in which groups of, or even individual,
columns would be read out through individual readout nodes.

The hit densities are predicted to be significantly 
lower at larger radius, and are not expected to be a concern
for layers 2-5, which lie at $\geq$24 mm from the beamline. 
If the inner layer were omitted, 
or if the whole detector were pushed out in radius to start
at 24 mm, the
extrapolation from the first track hit back towards the IP
would be doubled, and our simulations have shown that 
the flavour tagging performance would be noticeably worse.
Moreover, the backgrounds in the real accelerator may be
larger than the current estimates suggest, so that in any
case an increased CCD readout rate will help to secure
additional `headroom' against such an eventuality.

It is likely that by a combination of increased pixel readout 
rate and multiplexed CCD readout, a large detector B-field, 
and in the last resort
a (compromised) larger inner layer radius, the hit density
can be kept at or below the 10 hits/mm$^2$ level. 

\subsection{Radiation Damage Studies} 

Preliminary simulations~\cite{napoly} have indicated that the neutron
flux in the inner detector may be at the level of $10^8$ - 10$^9$ per
cm$^2$. Though orders of magnitude less than at the LHC, the rate is
large enough that more detailed simulations are warranted, and that
consideration be given to the radiation tolerance of the CCDs.

In many years of normal operating conditions at SLC, no radiation
damage was observed in the CCDs. However, during one unusual period
in which undamped beams were delivered for accelerator studies 
a noticeable charge-transfer inefficiency (CTI) was observed. This
effect was completely ameliorated by cooling the CCDs by an
additional 20 degrees to around 185~K.

Neutrons are believed to cause a CTI by producing bulk damage sites in
the silicon lattice, which act as charge-trapping centres. The 
minimum-ionising signal of roughly 2,000 electrons typically undergoes
several thousand serial transfers from pixel to pixel before reaching
the readout node, and a CTI of $\geq5\times10^{-4}$ would cause
a serious loss of signal. In recent neutron irradiation studies~\cite{sinev}
using VXD3 CCDs,
corresponding to an integrated dose of 6.5$\times10^9$ 1-4 MeV neutrons/cm$^2$,
a mean m.i.p. signal loss of 29\% was observed. Interestingly, after flushing
the CCD with charge, the loss was reduced to 18\%, and reduced further to
11\% by lowering the operating temperature from 185 K to 178 K.

These studies suggest that charge-flushing may serve to fill, at
least temporarily, the charge traps caused by radiation damage, and
that the full-trap lifetime can be extended by lowering the
temperature. We intend to pursue both approaches and there is
good reason to believe that CCDs can be made radiation tolerant
at the 10$^{10}$ neutrons/cm$^2$ level, which is 1-2 orders of
magnitude above the estimated flux at the LC. In addition,
the development of more optimised shielding strategies may serve
to reduce the expected neutron flux in the interaction region.
Finally, ideas to reengineer the CCD architecture, so as to 
reduce the effective charge-storage volume and hence the sensitivity 
to bulk damage, may also be pursued.

\section{Summary and Outlook}

In summary, CCDs offer a very attractive option for a high-energy
linear collider vertex detector. CCD VXDs have been `combat-tested'
at the first linear collider, SLC, and have allowed SLD to achieve
unrivalled $b$ and $c$-jet tagging performance.
Through further improvements, via the production of thinner,
larger-area, faster-readout CCDs, there is every reason to expect
that 21st-Century flavour-tagging at the next-generation linear 
collider will be substantially better, and able to meet the
demands of high-efficiency tagging in a multijet environment.

Groups centred in the USA/Japan and Europe are currently 
preparing the technical design reports for the accelerator
and detector(s), which will be presented to the respective funding
agencies in 2001/2. From the technical point-of-view construction could 
start as early as 2003, with first physics in 2008/9. 

The LCFI
Collaboration has started an R\&D programme to address the CCD
design issues, such that a vertex detector blueprint could be
credibly produced on a matching timescale. We look forward to
presenting the first results of this endeavour at Vertex2000!


\begin{thebibliography}{99}

\bibitem{accmor}
C.J.S. Damerell \etal, IEEE Trans. Nucl. Sci. {\bf 33} (1986) 51.\\
S. Barlag \etal, Phys. Lett. {\bf B184} (1987) 283.

\bibitem{toshi}
T. Abe, these proceedings.

\bibitem{vxd2}
C.J.S. Damerell \etal, Nucl. Instr. Meth. {\bf A288} (1990) 236.

\bibitem{vxd3}
K. Abe \etal, Nucl. Instr. Meth. {\bf A400} (1997) 287.
 
\bibitem{sldtag}
See \eg, SLD Collab., K. Abe \etal, Phys. Rev. {\bf D59} 052001 (1999).

\bibitem{dave}
D.J. Jackson, Nucl. Instr. Meth. {\bf A388} (1997) 247.

\bibitem{snowmass}
C.J.S. Damerell, D.J. Jackson, Proceedings of Snowmass 96 Workshop, p. 442.

\bibitem{lcfi}
LCFI (UK) Collaboration, `A proposal to initiate Research and Development for
a Vertex Detector at the future \ep collider', S.F.~Biagi \etal:\\
http://hep.ph.liv.ac.uk/\~~green/lcfi/lcfihome.html.

\bibitem{napoly}
O. Napoly, talk at LCWS99, Sitges, Spain: \\
http://www.cern.ch/Physics/LCWS99/talks.html.

\bibitem{sinev}
N. Sinev, talk at LCWS99, Sitges, Spain: \\
http://www.cern.ch/Physics/LCWS99/talks.html

\end{thebibliography}
\end{document}